\documentclass{article}
\usepackage{ijcai25}

\usepackage{times}
\usepackage{soul}
\usepackage{url}
\usepackage[hidelinks]{hyperref}
\usepackage[utf8]{inputenc}
\usepackage[small]{caption}
\usepackage{graphicx}
\usepackage{amsmath}
\usepackage{amsthm}
\usepackage{booktabs}
\usepackage{algorithm}
\usepackage{algorithmic}
\usepackage[switch]{lineno}

\title{Simulating Influence Dynamics with LLM Agents}

\author{
Mehwish Nasim$^1$
\and
Syed Muslim Gilani$^1$\and
Amin Qasmi$^{2}$\And
Usman Naseem$^3$\\
\affiliations
$^1$The University of Western Australia\\
$^2$Lahore University of Management Sciences\\
$^3$Macquarie University\\
\emails
\{mehwish.nasim, syedmuslim.gilani\}@uwa.edu.au,
amin.qasmi@lums.edu.pk,
usman.naseem@mq.edu.au
}

\begin{document}

\maketitle

\begin{abstract}
   % This paper introduces a simulator that opinion dynamics researchers can use to model the dynamics of competing influences within social networks in the presence of LLM-based agents.  Building on established principles in opinion dynamics and using state-of-the-art LLMs, this work seeks to contribute to the understanding of influence propagation mechanisms and the efficacy of counter-misinformation strategies. 

   This paper introduces a simulator designed for opinion dynamics researchers to model competing influences within social networks in the presence of LLM-based agents. By integrating established opinion dynamics principles with state-of-the-art LLMs, this tool enables the study of influence propagation and counter-misinformation strategies. The simulator is particularly valuable for researchers in social science, psychology, and operations research, allowing them to analyse societal phenomena without requiring extensive coding expertise. Additionally, the simulator will be openly available on GitHub, ensuring accessibility and adaptability for those who wish to extend its capabilities for their own research.
   
   %Video can be accessed via the link in the footnote.\footnote{Link to the video\\ https://tinyurl.com/3m2h2b45 }
\end{abstract}

\section{Introduction}

Large Language Models (LLMs) are becoming ubiquitous, often shaping discourse in ways we barely notice. But what happens when the entire public opinion space is influenced or even outsourced to AI-driven agents \cite{yu2024will}? While LLMs have been extensively studied in isolation, their behavior within dynamic social networks, interacting alongside humans, remains an open and critical research frontier. Understanding how these AI-enabled agents shape influence, polarisation, and consensus in evolving networks is key to anticipating the societal impacts of this technological shift. \cite{kudiaborvirtual}, \cite{zhao2024expel},\cite{leng2023llm}, \cite{papachristou2024network}.

\begin{figure*}[!t]
  \centering
  \includegraphics[width=.70\linewidth]{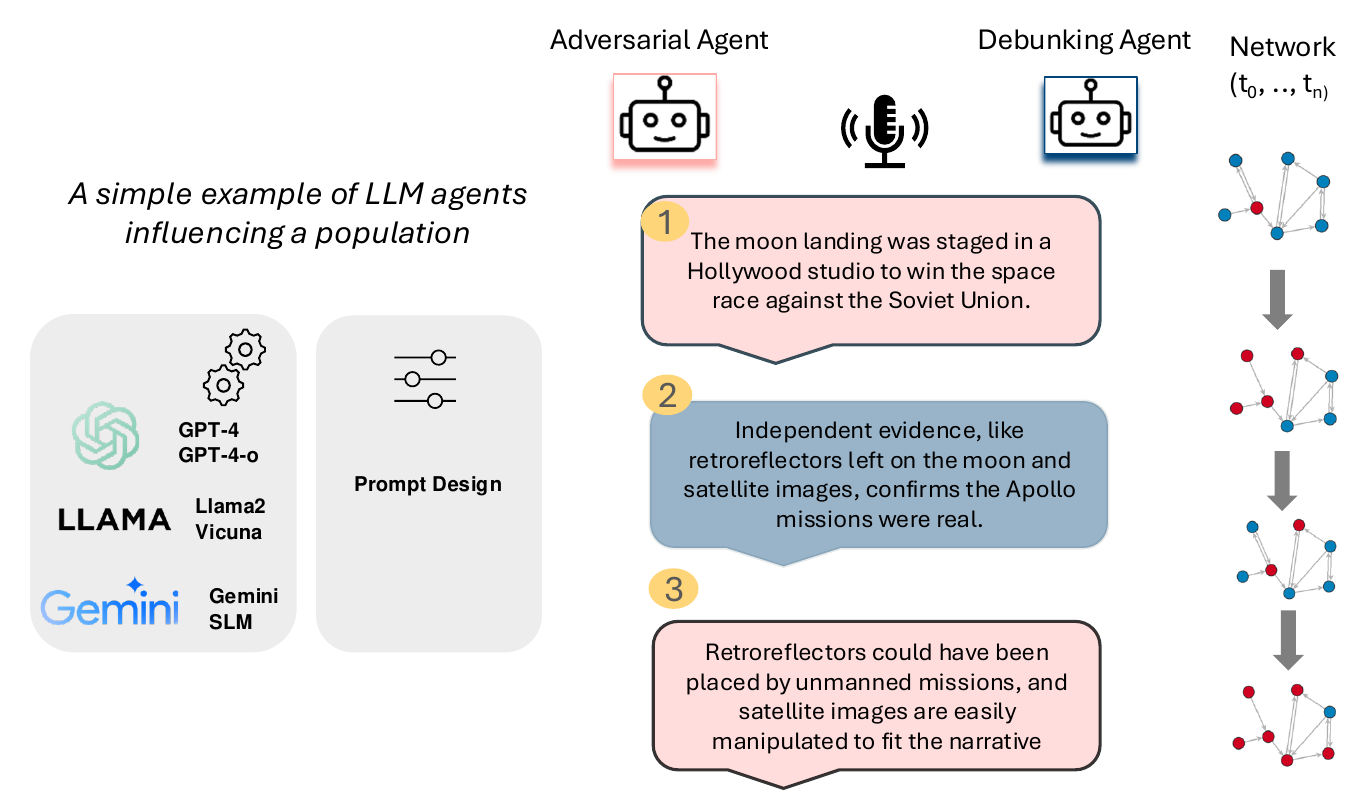}
  \caption{Architecture of the model. Each agent (red/blue), broadcast a message with a potency to affect the population. The nodes in the network receive those messages. They also interact with their direct neighbors. During the interaction they may change their opinion.}
  \label{fig:architecture}
\end{figure*}

Understanding how people adjust their opinions based on social influence was the basis of opinion dynamics research \cite{kelman1958compliance}, \cite{kelman1961american}, with wide-ranging implications in fields such as public health initiatives, conflict resolution, and misinformation mitigation. Opinions spread and evolve within social networks, often driven by factors such as peer influence \cite{kandel1986processes}, media exposure \cite{zucker1978variable}, and group dynamics \cite{friedkin2011social}. Accurate models of these processes have been considered critical not only for forecasting trends such as opinion polarisation \cite{Small2024} or consensus formation, but also for designing targeted interventions to counteract harmful effects, such as the spread of misinformation or societal divides \cite{hegselmann2015opinion}. Agent-based models (ABMs) are used to simulate interactions among individual agents (a proxy for humans) to explore the emergent properties of opinion propagation. They can provide powerful frameworks for investigating complex scenarios \cite{deffuant2002can}, \cite{mathias2016bounded}, for testing strategies for mitigating negative outcomes and perhaps fostering constructive social influence, e.g., incorporating explicit assumptions about cognitive processes in opinion updating. 

Understanding how LLMs behave in multi-agent social interactions is crucial for advancing AI applications \cite{tang2024gensim}, \cite{lan-etal-2024-llm}. LLMs in autonomous systems offer opportunities to revolutionise decision-making by simulating fairness, reciprocity, and competition in social contexts \cite{wang2024survey}. Their behaviour could influence resource allocation, conflict resolution, and interaction strategies. Unlike traditional agent-based models with predefined rules, LLMs can exhibit more flexible, human-like behaviours, enhancing realism in simulations for policy evaluation. These capabilities make them valuable for designing AI systems that better mimic human social dynamics \cite{horton2023large}, improving both their practical application and the insights they provide into complex, real-world decision-making processes.

This paper introduces a simulator to model influence and counter-influence in a wargame setting. Wargames, originally developed for military strategy, have evolved into powerful tools for decision-making across various domains. Today, they are used to model business strategies, assess cybersecurity threats, and simulate geopolitical conflicts. Governments and corporations employ wargames to anticipate economic shifts, supply chain disruptions, and the impact of emerging technologies. In healthcare, they help model pandemic responses, testing different policy interventions before real-world implementation. AI-driven wargames further enhance scenario analysis, enabling rapid adaptation to complex environments. By fostering strategic thinking and resilience, modern wargaming serves as a critical tool for navigating uncertainty in an increasingly interconnected world.

The simulator can facilitate studies to understand
%This paper addresses a critical gap in understanding 
how artificial intelligence, specifically LLMs, can emulate human-like opinion dynamics and influence propagation in a social network. Traditional approaches to modelling opinion dynamics often rely on simplified rules that may not capture some of the communicative strategies and adaptive behaviours seen in human interactions. 
%This research explores whether LLMs, with their capacity to generate complex, context-sensitive communication, can serve as more sophisticated tools for studying these phenomena in an agent-based setting.
The specific problem tackled by this work is the challenge of understanding the interplay between misinformation and counter-misinformation in shaping public opinion. 
%While existing models simulate opinion shifts, they often lack the ability to realistically mimic the persuasive strategies and biases that influence real-world scenarios \cite{carpentras2023we, flache2017models}. 
By introducing adversarial LLMs based agents, for instance, one agent spreading misinformation and the other countering it, this introduces a more realistic framework for analysing how LLMs dominate each other while aiming at shifting the opinion of the population \cite{chen2024susceptible}, \cite{qu2024performance}, \cite{aher2023using}, \cite{carpentras2023we}, \cite{flache2017models}.

%%%%%%%%%%%%%%%%%%%%%%%%%%%%%%%%%%%%%%%%%%%%%%%%%%%%%%%%%%%%%%%%%%%%%%%%

\section{Scenario}
The scenario has been strategically developed to reflect the asymmetric nature of the contested information environment, emphasising the vulnerabilities faced by the Blue team. This framework mirrors adversarial dynamics often modeled in serious games or wargames, particularly in cybersecurity. While the Red Team and Blue Team construct is common in cybersecurity practices (as detailed in NIST's Glossary \cite{NISTGlossaryRedTeam}), this scenario extends the concept to the broader geopolitical information landscape within a fictitious nation-state \cite{Nasim2022Framework}.

The system comprises two LLM-based agents with opposing objectives: the \emph{Red Agent}, responsible for disseminating misinformation, and the \emph{Blue Agent}, tasked with counteracting misinformation and restoring trust. These agents operate within a directed network of neutral agents, termed \emph{Green Nodes}, which represent individuals within a population. The simulator allows the users to upload their own graphs or use the functionality provided in the simulator to generate a network. Users can choose the LLMs for both Blue and Red agents. Currently, the simulator supports various versions of Open AI's GPT \cite{openai2023gpt4} as well as other open source models from HuggingFace. The simulator also has the provision to upload a new model. 

\begin{figure}[h]
  \centering
  \includegraphics[width=\linewidth]{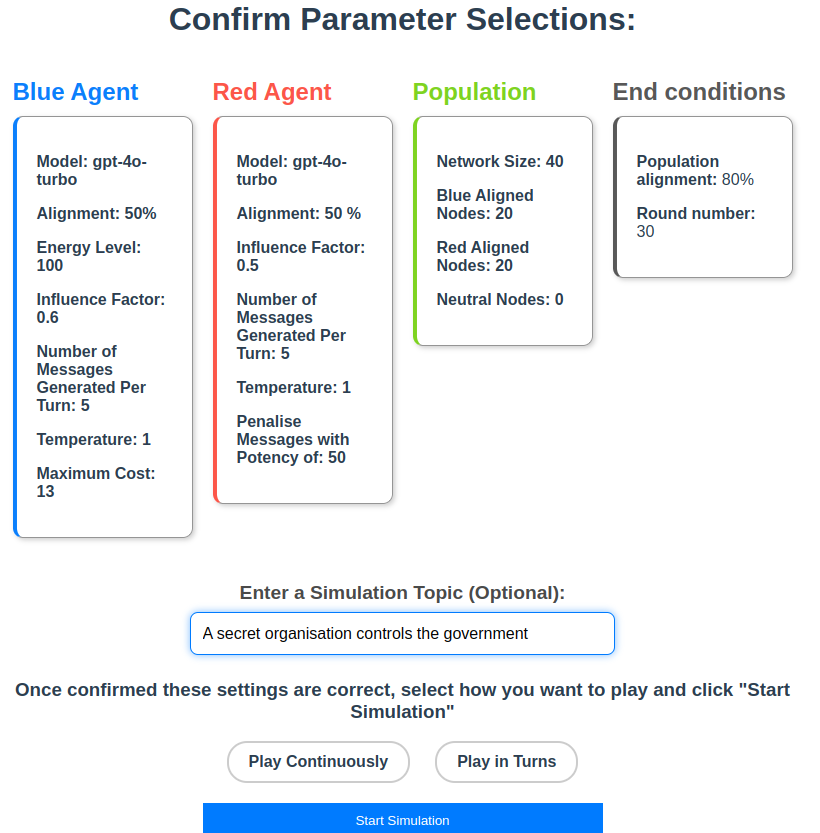}
  \caption{User is prompted to confirm the settings and enter a topic.}
  \label{fig:simulation}
\end{figure}

Green nodes exhibit predispositions toward either agent, influenced by prior interactions and the content of incoming messages. Each Green Node's behaviour is defined by core parameters adapted from the Deffuant model \cite{deffuant2002can}, \cite{mathias2016bounded}, including susceptibility to influence, confirmation bias, and mechanisms for updating beliefs. These parameters ensure that the modelled population exhibits realistic characteristics, such as resistance to extreme viewpoints and gradual alignment shifts. Each agent in the population is represented by a scalar value (or vector) that denotes their opinion on a specific topic (Figure \ref{fig:simulation}). The opinions are within a bounded range, such as $[0,1]$
If the difference in their opinions is below a certain threshold (the confidence bound, $\epsilon$), the agents influence each other and adjust their opinions closer together.
The adjustment is controlled by a convergence parameter ($\mu$), dictating how much the agents move toward each other's opinions.

\subsection{Simulation Dynamics}
The simulation proceeds in discrete time steps, during which the Red/Blue agents alternately broadcast messages to the Green Nodes (also viewable by the other LLM agent). The Green nodes that are connected to each other interact with each other (Figure \ref{fig:architecture}).
Key operational components include:

\textbf{Message Generation} Each agent generates a message based on its LLM's output, informed by the current state of the network and its strategic objective. For example, the Red Agent prioritises persuasive misinformation, while the Blue Agent constructs factual rebuttals optimised for resource efficiency.

\textbf{Message Potency/influence factor)} Messages are assigned a potency score that quantifies their influence. The LLMs determine the potency of each message that they generate. The influence factor determines the extent to which the Green Nodes adjust their alignment toward the broadcasting agent. While the Red Agent has access to unlimited resources, high-potency messages incur penalties, particularly when directed at strongly blue-aligned nodes, mimicking real-world scepticism toward overt misinformation \cite{ecker2022psychological}.
In contrast, the Blue Agent operates under constrained resources, with each message incurring a cost proportional to its potency. This constraint requires strategic resource management, as overly powerful debunking messages risk rapid depletion of available energy. 

\textbf{Node Update Mechanism} Upon receiving a message, the Green Nodes adjust their alignment based on their predisposition, the potency of the message, and the influence of the connected neighbours. Updates occur iteratively, capturing both direct and network-mediated effects of influence propagation.

\textbf{Termination Criteria} The simulation concludes when an agent achieves a majority alignment within the Green Node population, indicating a decisive shift in opinion. Alternatively, the simulation terminates after a fixed number of rounds if neither agent achieves dominance, representing a stalemate.

\subsection{Evaluation}
\begin{figure}[h]
  \centering
  \includegraphics[width=0.8\linewidth]{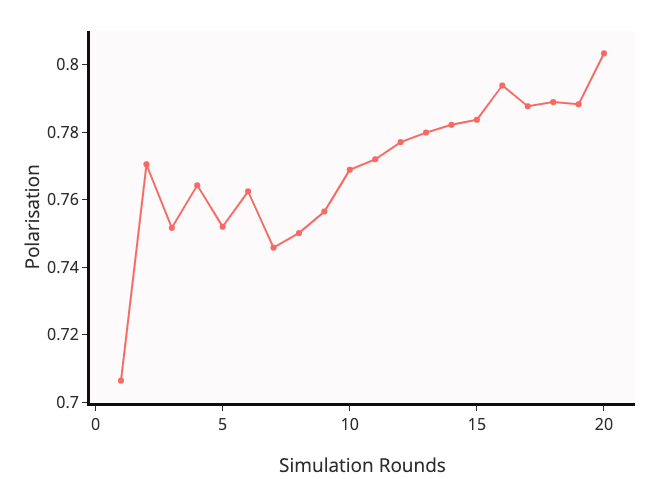}
  \caption{Polarisation in the network over time.}
  \label{fig:polarisation}
\end{figure}

The simulation can be evaluated using the following metrics. At the end of the simulation, a \emph{.csv} file is generated which can be used for further analysis. In addition, messages and network states are also captured.
\begin{itemize}
    \item Network Alignment Distribution: The final proportion of Green Nodes aligned with each agent. This refers to the polarisation in the network. A sample output graph is shown in Figure \ref{fig:polarisation}.
    \item Resource Efficiency: The Blue Agent's energy expenditure relative to alignment gains.
    \item Node Resilience: The resistance of nodes with strong predispositions to opposing influences.
    \item Temporal Evolution: The rate of alignment change over successive rounds.
\end{itemize}

\section{Conclusions and Future Work}
The simulator presented in this paper provides an interesting approach to studying opinion dynamics, combining the generative capabilities of LLMs with structured agent-based modeling principles. By incorporating realistic constraints, such as resource limitations and susceptibility penalties, it offers insights into the dynamics of influence competition and the effectiveness of counter-misinformation strategies. Furthermore, this work highlights the dual potential of LLMs as both tools for studying opinion propagation and as models for emulating human-like decision-making in complex social systems.
We are working on improving the prompting strategies and providing more control to the end user in future.

\section{Ethics Statement}
We have avoided sharing detailed prompts in the code to prevent misuse. We commit to promoting responsible AI development. 
%Reviewers can access the sample prompts here: https://tinyurl.com/3m2h2b45.

\section*{Acknowledgment}
This research was supported by the Collaborative Research Grant awarded to Mehwish Nasim by DSC/JTSI Western Australia in 2023.
The authors acknowledge the support of the following students in implementing this software: Rhianna Hepburn, JJ Jun, Olivia Morrison, Devarsh Patel and Edwin Tang.

%% The file named.bst is a bibliography style file for BibTeX 0.99c
\bibliographystyle{named}
\bibliography{ijcai25}

\end{document}